\documentclass[twocolumn,english]{revtex4-2}
\usepackage[T1]{fontenc}
\usepackage[latin9]{inputenc}
\usepackage{babel}
\usepackage{float}
\usepackage{amstext}
\usepackage{amssymb}
\usepackage{graphicx}
\usepackage{wasysym}
\usepackage[unicode=true,pdfusetitle,
 bookmarks=true,bookmarksnumbered=false,bookmarksopen=false,
 breaklinks=false,pdfborder={0 0 1},backref=false,colorlinks=false]
 {hyperref}

\makeatletter

\providecommand{\tabularnewline}{\\}

\makeatother

\begin{document}
\title{Ising model on a 2D \emph{additive} Small-World Network}
\author{R. A. Dumer }
\email{rafaeldumer@fisica.ufmt.br}

\affiliation{Instituto de Física - Universidade Federal de Mato Grosso, 78060-900,
Cuiabá, Mato Grosso, Brazil.}
\author{M. Godoy}
\email{mgodoy@fisica.ufmt.br}

\affiliation{Instituto de Física - Universidade Federal de Mato Grosso, 78060-900,
Cuiabá, Mato Grosso, Brazil.}
\begin{abstract}
In this article, we have employed Monte Carlo simulations to study
the Ising model on a two-dimensional \emph{additive} small-world network
(A-SWN). The system model consists of a $L\times L$ square lattice
where each site of the lattice is occupied for a spin variable that
interacts with the nearest neighbor and has a certain probability
$p$ of being additionally connected at random to one of its farther
neighbors. The system is in contact with a heat bath at a given temperature
$T$ and it is simulated by one-spin flip according to the Metropolis
prescription. We have calculated the thermodynamic quantities of the
system, such as, the magnetization per spin $m_{L}$, magnetic susceptibility
$\chi_{L}$, and the reduced fourth-order Binder cumulant $U_{L}$
as a function of $T$ for several values of lattice size $L$ and
additive probability $p$. We also have constructed the phase diagram
for the equilibrium states of the model in the plane $T$ versus $p$
showing the existence of a continuous transition line between the
ferromagnetic $F$ and paramagnetic $P$ phases. Using the finite-size
scaling (FSS) theory, we have obtained the critical exponents for
the system, where varying the parameter $p$, we have observed a change
in the critical behavior from the regular square lattice Ising model
to A-SWN.
\end{abstract}
\pacs{9999.nt}
\maketitle

\section{Introduction}

At the beginning of the century, the Small-World effect observed by
Milgram \citep{1} came to prominence after Watts and Strogatz \citep{2}
quantify their structural properties through graph theory. The Watts-Strogatz
model \citep{2} (WS-model) is based on the idea that the vertices
of a graph are the network sites, and the edges are connections between
two sites of this network. As well, by introducing a disorder parameter
$p$, as the probability of rewriting randomly each one of the connections
in a regular lattice, we obtain the small-world network (SWN) in specific
regions of the interval $0<p\leq1$. These regions of $p$ in which
we have the SWN regime, it is identified when the network has the
local clustering of a regular lattice, but at the same time have an
average distance between any two sites of the network characteristic
of a random lattice \citep{2}.

The structural properties of the WS-model were investigated \citep{2,3,4,5,6,7}
and applied in the study of the disease transmission with the probability
reaching the epidemic behavior \citep{2,4,8}, once in the Milgram
experiment \citep{1} is suggested that on average, exist six intermediaries
separating two people in the world, inducing that one high infectious
disease could spread across the whole planet in about six incubation
periods for the disease \citep{8}.

In addition to the WS-model, some variants of this model were also
developed to describe the properties of an SWN. In one of these variants
\citep{4}, there is a regular square lattice where each site has
$n$ edges, i.e., degree $n$, and with a certain probability $p$
is possible to add long-range interactions to each site in the network,
being each site able to receive $n$ long-range interactions. Therefore,
we can obtain a small typical separation, while the clustering property
of the regular lattice is always preserved. A simplified version of
this variant can be made if we decrease the number of long-range interactions
that each site can receive. These variants are called \emph{additive}
SWN (A-SWN), while we have the in the WS-model we have defined the
\emph{rewiring} SWN (R-SWN) \citep{2}.

Since the first SWN model was proposed, these networks have been implemented
in a variety of physical models \citep{9,10,11,12}, including the
Ising model, which was studied by Monte Carlo simulations with the
A-SWN in 1D \citep{13,14} and with the R-SWN in 1D, 2D e 3D \citep{15,16}.
We also have exact and approximate analytical results for the A-SWN
and R-SWN in 1D \citep{5,17,18}. These results in 1D show that an
ordered to the disordered phase transition is obtained for $T\ne0$
with $0<p\le1$, and for 1D, 2D e 3D is observed a change in the critical
behavior of the system by the addition of long-range interactions.

In this work, we have investigated the Ising model in a two-dimensional
A-SWN, where each site of the network is occupied by a spin variable
$\sigma=1/2$ that can assume values $\pm1$. We limit by one the
number of long-range interactions that each site can receive with
probability $p$ and dividing our network in two sublattices, each
new interaction created should connect these sublattices. A similar
model was proposed by Zhang and Novotny \citep{19} in his SW-model,
where the long-range interactions are completely random and always
present on all sites of the network, i.e., $p$ is always $1$. They
have found a mean-field critical behavior in his model, so here we
will verify if the limitation on the randomness of the long-range
interactions changes the predicted mean-field critical behavior for
$p=1$, and what is the critical behavior of the system in other points
of the A-SWN regime, i.e., other regions of $p$ where the A-SWN behavior
is also observed.

This article is organized as follows: In Section \ref{sec:Model},
we describe the network used and the Hamiltonian model of the system.
In Section \ref{sec:Monte-Carlo-simulations} we present the Monte
Carlo simulation method used, some details concerning the simulation
procedures, and the thermodynamic quantities of the system necessary
for the application of FSS analysis. The behavior of the phase transitions,
phase diagrams, and critical exponents by FSS analysis are described
in Section \ref{sec:Results}, and finally, in Section \ref{sec:Conclusions}
we present our conclusions.

\section{Model\label{sec:Model}}

\begin{figure}
\begin{centering}
\includegraphics[clip,scale=0.73]{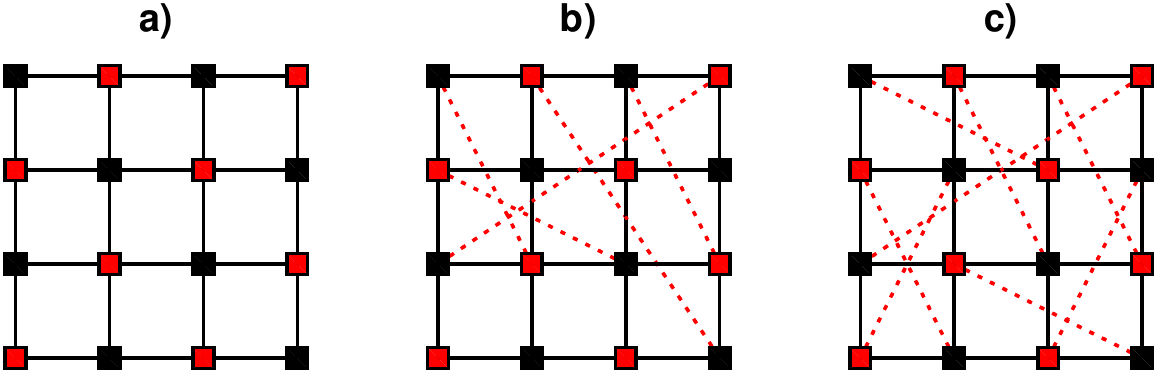}
\par\end{centering}
\caption{{\footnotesize{}Schematic representation of the system and the A-SWN.
The black square dots indicate the sites on one of the sublattices,
the red square dots are the sites on the other sublattice, the solid
black lines are the nearest-neighbor interactions $J$ between pairs
of spins, and the red-dashed lines are long-range interaction} $J_{ik}${\footnotesize{}
added to the network with a certain probability $p$. In a) we have
$p=0$, i.e., the probability of adding a long-range interaction}
$J_{ik}${\footnotesize{} to any site on the lattice is zero, therefore,
we have a regular square lattice. b) $p=0.5$, we are in the A-SWN
regime because in addition to the conservation of $C(p)$, and we
also have an average the short path length between network sites,
through the shortcuts created by the long-range interaction} $J_{ik}${\footnotesize{}
added between the sublattices. c) $p=1$, all sites on the network
have a long-range interaction} $J_{ik}${\footnotesize{} connecting
the two sublattices, and consequently, it is the network with the
shortest typical separation between the sites on the network.\label{fig:1}}}
\end{figure}

In this work, the studied model is the Ising model with $N=L^{2}$
spins $\sigma_{i}=\pm1$ on a regular square lattice $L\times L$,
with periodic boundary conditions, and a nearest-neighbor ferromagnetic
interaction of strength $J$ (see Fig. \ref{fig:1}(a)). On the other
hand, using the same regular square lattice $N=L\times L$, with a
certain probability $p$, we can add one long-range interaction $J_{ik}$
to each site of the lattice. To add the long-range interaction $J_{ik}$,
we divide the system in two sublattices, where one sublattice plays
the role of central spins, while the other sublattice contains the
spins in which the central spins can connect, beyond the nearest neighbors.
Thus, to choose a long-range interaction $J_{ik}$ for a site $i$,
the sublattice of $i$ will be the sublattice of the central spins,
then, we choose randomly a site $k$ from another sublattice. If the
site $k$ does not be one of its nearest neighbors already naturally
coupled with $i$, we picked a random number $0<r<1$, and if $r\le p$
(with $p$ predefined), we couple the site $k$ to the neighbors of
the site $i$, and for the site $k$ we couple the site $i$ to its
neighbors. The try to add a long-range interaction $J_{ik}$ is made
once to each site that do not have a long-range interaction $J_{ik}$
in the network, and as result we have a network with average coordination
number $z=4+p$.

We can see the evolution of the system for some values of $p\neq0$
(see Fig. \ref{fig:1}(b) and (c)). To reach the A-SWN state as defined
in WS-model with the R-SWN, we need to have two structural properties
on the network: 1) a high clustering coefficient $C(p)$, which is
characteristic of regular lattices and is defined as the ratio between
the number of connections among neighbors at any site of the network
and the possible number of connections among this quantity of neighbors.
2) a small characteristic length path $l(p)$, which is observed in
random networks and defined as the shortest distance between two any
sites of the network.

Thus, as our regular structure in $p=0$ keeps unaltered, we have
a high $C(p)$ for any value of $p$, and conform we increase $p$,
the long-range interaction $J_{ik}$ is added to the network, creating
shortcuts between the sites that before in the simple regular lattice
would be more distant, consequently decreasing the $l(p)$ of the
network. The $l(p)$ scales linearly $l(p\to0)\sim L/2$, and logarithmically
$l(p\to1)\sim\ln(L^{1.77})$, being these regimes referred as the
``large-world'' and ``small-world'' respectively. The cross-over
between these regimes occurs when the average number of shortcuts
is about one, or in the other words, we can say in the SWN regime
when $p\apprge L^{-2}$ \citep{4}. Versed on this, our study is based
on $p\ge0.25$ values, where the A-SWN is found and the decay of $l$
as a function of $p$ undergoes less, i.e., having approximately the
same value of $l$.

The ferromagnetic Ising spin energy is described by the Hamiltonian
of the form

\begin{equation}
{\cal H}=-J\sum_{\left\langle i,j\right\rangle }\sigma_{i}\sigma_{j}-\sum_{\left\langle i,k\right\rangle }J_{ik}\sigma_{i}\sigma_{k},\label{eq:1}
\end{equation}
where $J$ is the nearest-neighbor ferromagnetic interaction, $J_{ik}$
is the long-range interaction on the A-SWN. The first sum is over
all the pair of nearest-neighbor spins on the square regular lattice
and the second sum is made over all the pairs of spins $(i,k)$ connected
through long-range interaction on the A-SWN. 

The long-range interaction $J_{ik}$ is distribute randomly and satisfy
the following probability distributions: 
\begin{equation}
P(J_{ik})=(1-p)\delta(J_{ik}-0)+p\delta(J_{ik}-J),\label{eq:2}
\end{equation}
where the term $(1-p)\delta(J_{ik}-0)$ indicated that one fraction
$(1-p)$ of pairs of spins $(i,k)$ on the lattice are free of the
long-range interactions, while the terms $p\delta(J_{ik}-J)$ indicated
that one fraction $p$ of pairs of spins $(i,k)$ are connected through
a long-range interaction. Here, we always are considering $J_{ik}=J=1$.

\section{Monte Carlo simulations\label{sec:Monte-Carlo-simulations}}

We simulate the system specified by the Hamiltonian in Eq. (\ref{eq:1})
on a $L\times L$ square lattice under periodic boundary conditions
applied in all directions. We have chosen the initial state of the
system with all spins aligned in the same direction, and a new configuration
is generated by the following Markov process: for a given temperature
$T$ and an additive probability $p$, we choose a random spin $\sigma_{i}$
from the square lattice, and then run the one-spin flip dynamic. In
this dynamic the flipping probability is dependent on the transition
rate $W(\sigma_{i}\to\sigma_{i}^{\prime})$, which is given by the
Metropolis prescription as follows

\begin{equation}
W(\sigma_{i}\to\sigma_{i}^{\prime})=\left\{ \begin{array}{cccc}
e^{(-\Delta E/k_{B}T)} & \textrm{if} & \Delta E>0\\
1 & \textrm{if} & \Delta E\le0 & ,
\end{array}\right.\label{eq:3}
\end{equation}
where $\Delta E$ is the change in energy after flipping the spin,
$\sigma_{i}\to\sigma_{i}^{\prime}$, $k_{B}$ is the Boltzmann constant,
and $T$ the temperature of the system. The new state is accepted
if $\Delta E\le0$, in the case of $\Delta E>0$ we choose another
random number $1<\xi<0$ and if $\xi\le\exp(-\Delta E/k_{B}T)$ the
new state is also accepted, but if none of the conditions are satisfied,
we do not change the state of the system. Repeating the Markov process
$N$ times, we have one Monte Carlo Step (MCS). In our simulation,
we have waited for $2\times10^{4}$ MCS for the system to reach the
stationary state for all the lattice sizes. We used more $5\times10^{3}$
MCS to calculate the thermal averages of the quantities of interest.
The average over the samples was done using 25 independent samples
for any lattices.

The measured thermodynamic quantities in our simulations are: magnetization
per spin $m_{L}$, magnetic susceptibility $\chi_{L}$ and reduced
fourth-order Binder cumulant $U_{L}$:

\begin{equation}
m_{L}=\frac{1}{N}\left[\left\langle \sum_{i=1}^{N}\sigma_{i}\right\rangle \right],\label{eq:4}
\end{equation}

\begin{equation}
\chi_{L}=\frac{N}{k_{B}T}\left[\left\langle m_{L}^{2}\right\rangle -\left\langle m_{L}\right\rangle ^{2}\right],\label{eq:5}
\end{equation}

\begin{equation}
U_{L}=1-\frac{\left[\left\langle m_{L}^{4}\right\rangle \right]}{3\left[\left\langle m_{L}^{2}\right\rangle ^{2}\right]},\label{eq:6}
\end{equation}
 where $\left[\ldots\right]$ denotes the average over the $25$ samples
and $\left\langle \ldots\right\rangle $ is the thermal average over
the $5\times10^{3}$ MCS. The lattice sizes from $L=24$ to $L=256$
are simulates and the data are analyzed via the FSS theory. The above-defined
quantities obey the following FSS relations in the neighborhood of
the critical temperature $T_{c}$:

\begin{equation}
m_{L}=L^{-\beta/\nu}m_{0}(L^{1/\nu}\varepsilon),\label{eq:7}
\end{equation}

\begin{equation}
\mathcal{\chi}_{L}=L^{\gamma/\nu}\mathcal{\chi}_{0}(L^{1/\nu}\varepsilon),\label{eq:8}
\end{equation}

\begin{equation}
U_{L}=U_{0}(L^{1/\nu}\varepsilon),\label{eq:9}
\end{equation}
where $\epsilon=(T-T_{c})/T_{c}$, and $m_{0}(L^{1/\nu}\varepsilon)$,
$\chi_{0}(L^{1/\nu}\varepsilon)$ and $U_{0}(L^{1/\nu}\varepsilon)$
are scaling functions, and $\beta$, $\gamma$ and $\nu$ are the
magnetization, magnetic susceptibility and length correlation critical
exponents, respectively. The derivative of Eq. (\ref{eq:9}) with
respect to the parameter $T$ give us the following scaling relation:

\begin{equation}
U'_{L}=L^{1/\nu}\frac{U'_{0}(L^{1/\nu}\varepsilon)}{T_{c}}.\label{eq:10}
\end{equation}

We have determined the critical exponents $\beta/\nu$, $\gamma/\nu$
and $\nu$ from slope of a log-log plot of $m_{L}(T_{c})$, $\mathcal{\chi}_{L}(T_{c})$
or $U'_{L}(T_{c})$ versus lattice size $L$, respectively. We also
have used another alternative method to estimate the values of the
critical exponents, the data collapse from the scaling functions \citep{20,21,22}.

\section{Results\label{sec:Results}}

\begin{figure}
\begin{centering}
\includegraphics[clip,scale=0.55]{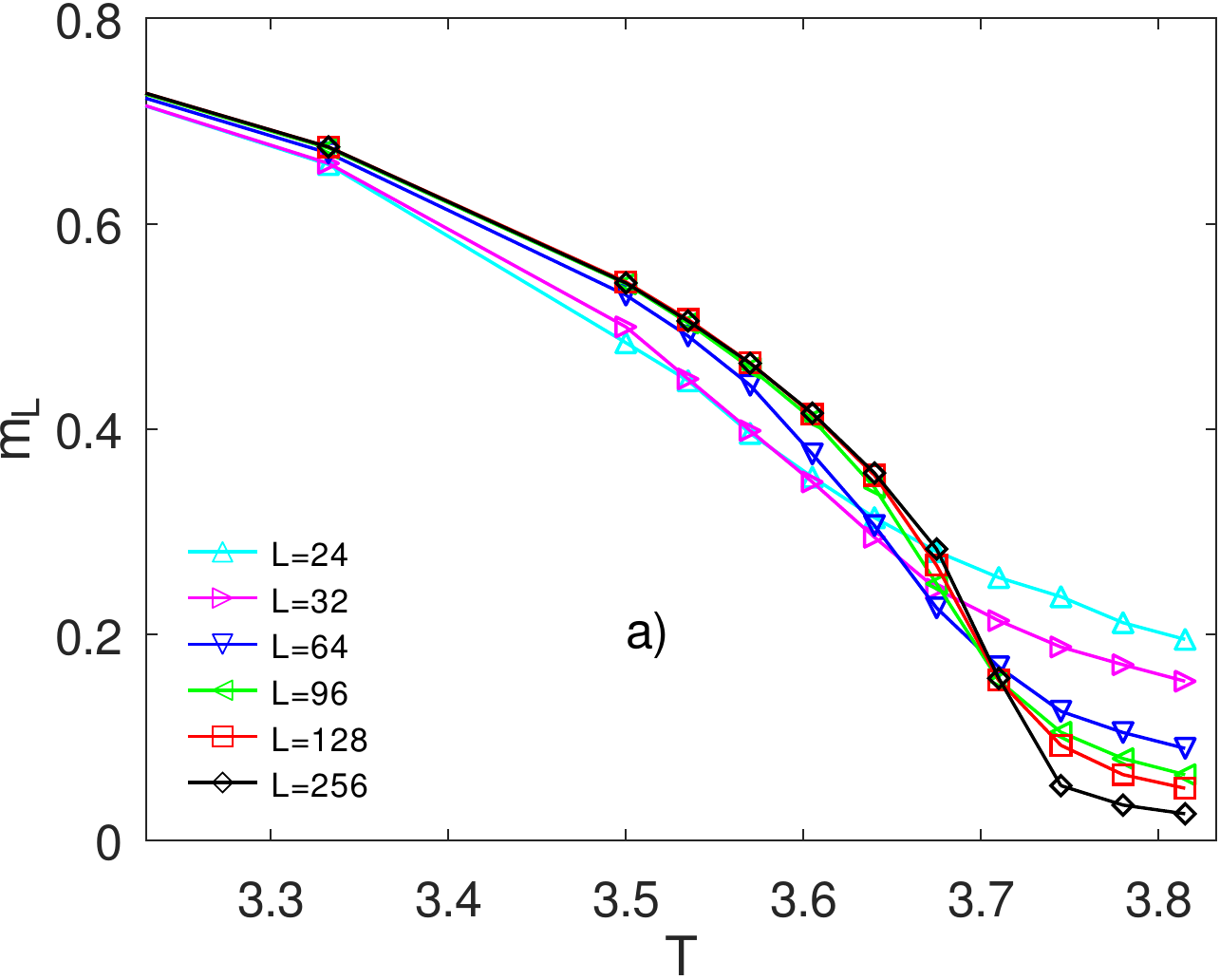}
\par\end{centering}
\begin{centering}
\includegraphics[clip,scale=0.55]{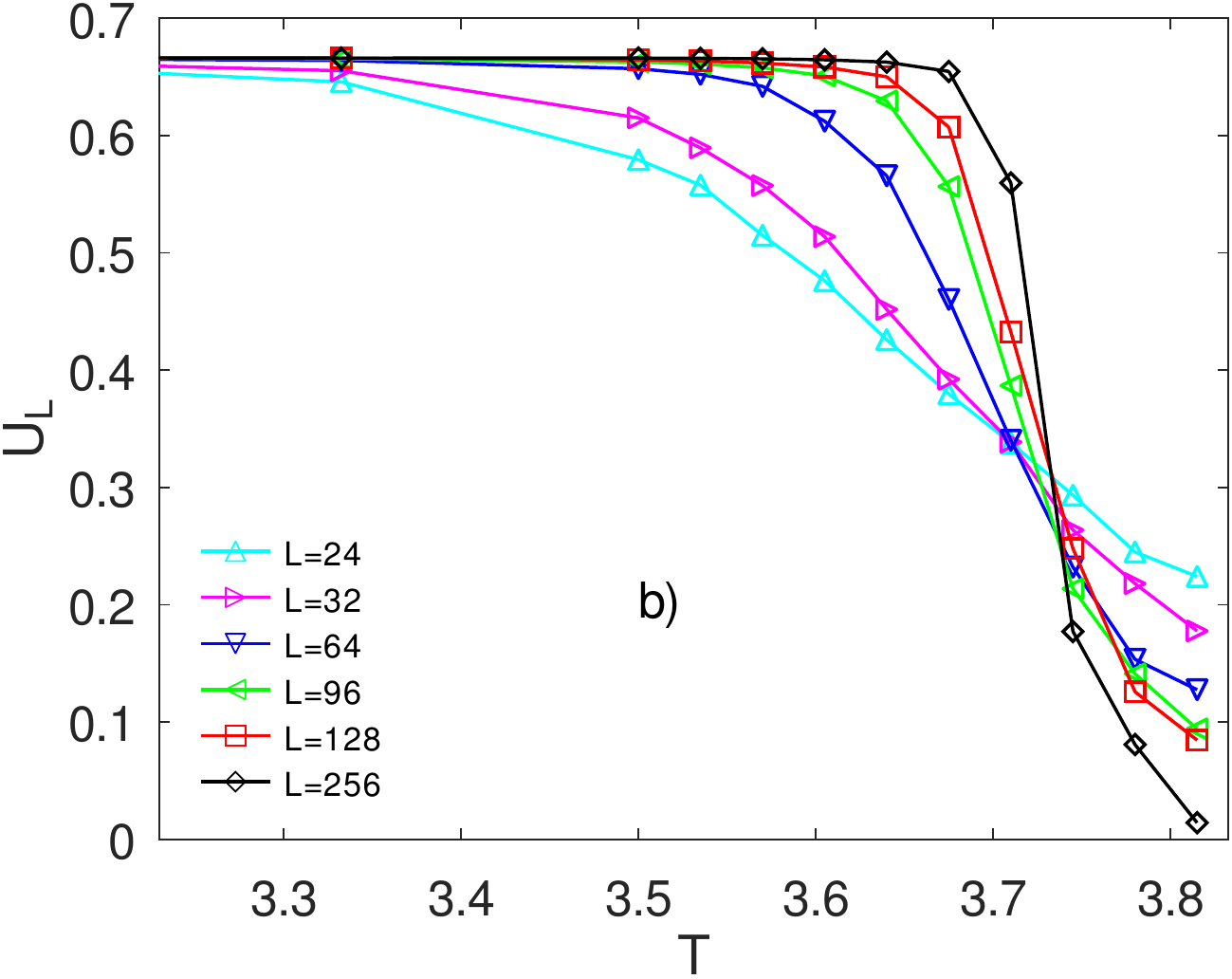}
\par\end{centering}
\begin{centering}
\includegraphics[clip,scale=0.55]{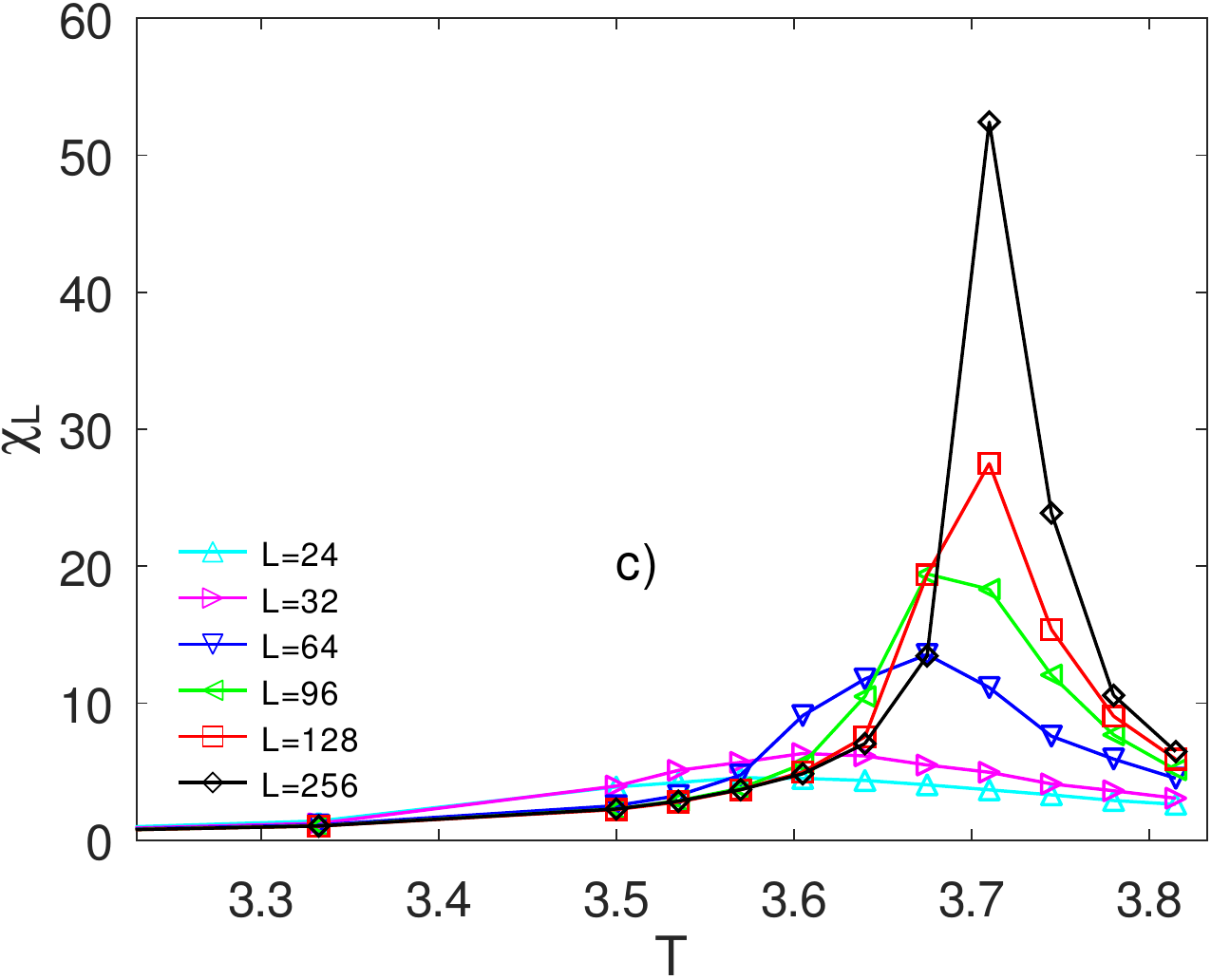}
\par\end{centering}
\caption{{\footnotesize{}Finite-size behavior of the a) magnetization $m_{L}$,
b) fourth-order Binder cumulant $U_{L}$, and c) magnetic susceptibility
$\chi_{L}$ as a function of temperature $T$ for several lattice
sizes $L,$ as indicated in the figures. Here, we have used a fixed
additive probability $p=0.75$. The error bars are within the symbol
size.\label{fig:2}}}
\end{figure}

In this section, we present the results for the magnetic properties
of the Ising model on a 2D A-SWN. The Ising model is very useful to
identify the phase transitions in magnetic systems. This can be made
by observing the behavior of the magnetic susceptibility due to its
discontinuously at the critical point and thermodynamic limit. On
the other hand, computationally this limit is impracticable, so we
use some techniques to study the critical behavior for finite-size
lattices \citep{20,21,22}. One of these techniques is to calculate
the temperature $T(\chi_{L}^{max})$ for the locations of the maxima
magnetic susceptibility peaks for each $L$ and to plot versus $L^{-1}$.
The critical temperature can be estimated from an infinite-size extrapolation
in according to $T(\chi_{L}^{max})-T_{c}(\infty)=\alpha L^{-1}$.
In addition, to here we also have used the crossing of the reduced
fourth-order Binder cumulant $U_{L}$ for different lattice sizes
$L$ to identify the critical temperature and the second-order phase
transition in the system \citep{23}. For $L\longrightarrow\infty$,
we have that $U_{\infty}\to2/3$ in the ordered phase, and $U_{\infty}\to0$
in the disordered phase. We observed a singular point independent
of the lattice sizes and correspond to the critical point of the phase
transitions \citep{20,21,22,23}. The critical temperature obtained
by this method is in agreement with those obtained from the maxima
of the magnetic susceptibility. 

\begin{figure}
\begin{centering}
\includegraphics[clip,scale=0.55]{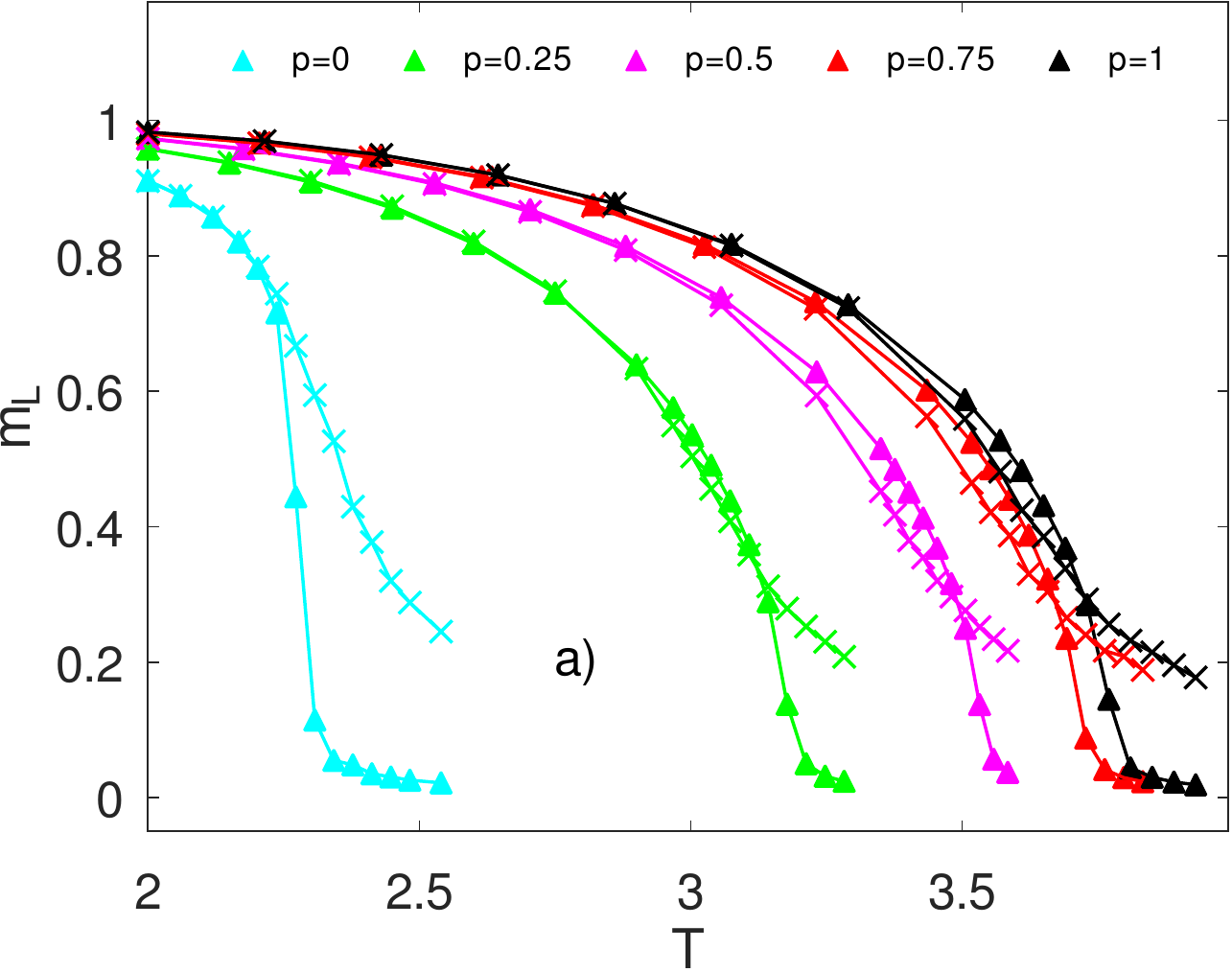}
\par\end{centering}
\begin{centering}
\includegraphics[clip,scale=0.55]{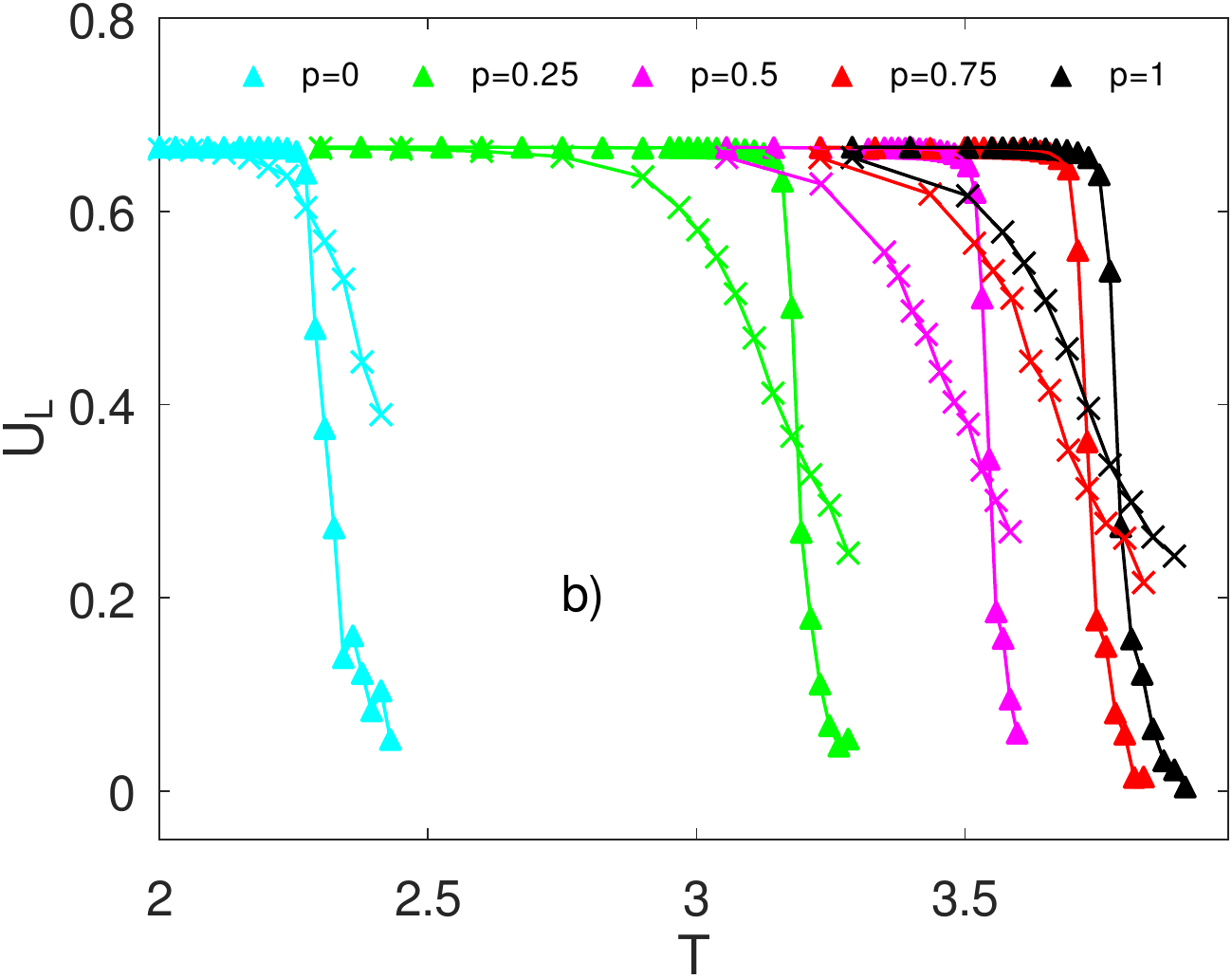}
\par\end{centering}
\begin{centering}
\includegraphics[clip,scale=0.55]{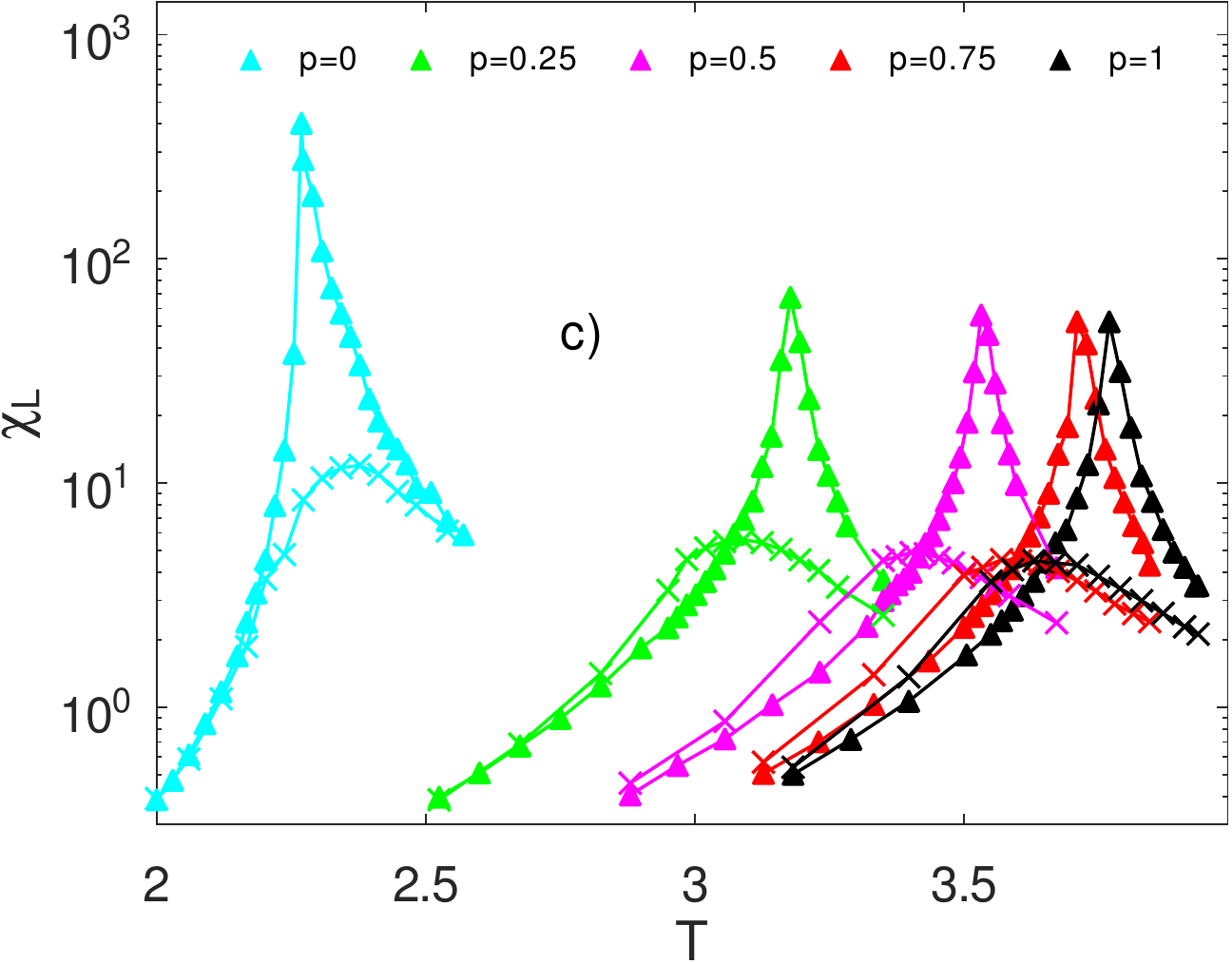}
\par\end{centering}
\caption{{\footnotesize{}a) Magnetization $m_{L}$, b) fourth-order Binder
cumulant $U_{L}$, and c) magnetic susceptibility $\chi_{L}$ as a
function of temperature $T$ for two different lattice sizes $L=256\:(\triangle)$
and $L=24\:(\times)$, and for several values of $p$, as indicated
in the figures. The error bars are within the symbol size.\label{fig:3}}}
\end{figure}

In the A-SWN regime ($0<p\leq1$), we have used $p=0.25$, $p=0.50$,
$p=0.75$ and $p=1$ to study the critical behavior of the system.
The best results are obtained from $p=1$ and $p=0.75$, where the
majority quantity of the sites contains the same coordination number
$z=5$. In Fig. \ref{fig:2}, we can see the finite-size behavior
of some thermodynamic quantities, such as the magnetization $m_{L}$,
the fourth-order Binder cumulant $U_{L}$, and the magnetic susceptibility
$\chi_{L}$ as a function of temperature $T$ and for several lattice
sizes $L,$ in the A-SWN regime with $p=0.75$. The finite-size behavior
is observed when the magnetization vanishes with increasing the temperature
$T$ (see Fig. \ref{fig:2}(a)) and the magnetic susceptibility presents
a peak around the critical temperature $T_{c}$, which grows in height
with the increase of $L$ indicating the existence of a phase transition
(see Fig. \ref{fig:2}(c)). The position of the magnetic susceptibility
peaks can be defined at a pseudo-critical temperature $T(\chi_{L}^{max})$.
The $T(\chi_{L}^{max})$ approaches $T_{c}(\infty)$ of the system
when $L\rightarrow\infty$. To study the phase transition in more
detail, we also used the fourth-order cumulants $U_{L}$ intersection
method to determine the value of temperature at which the transition
occurs. In order to find the critical temperature, we display in Fig.
\ref{fig:2}(b) the cumulants $U_{L}(T)$ vs temperature $T$ for
several system sizes $L$. For example, our estimate for the dimensionless
critical temperature is $T_{c}=3.73\pm0.02$. 

\begin{figure}
\begin{centering}
\includegraphics[clip,scale=0.55]{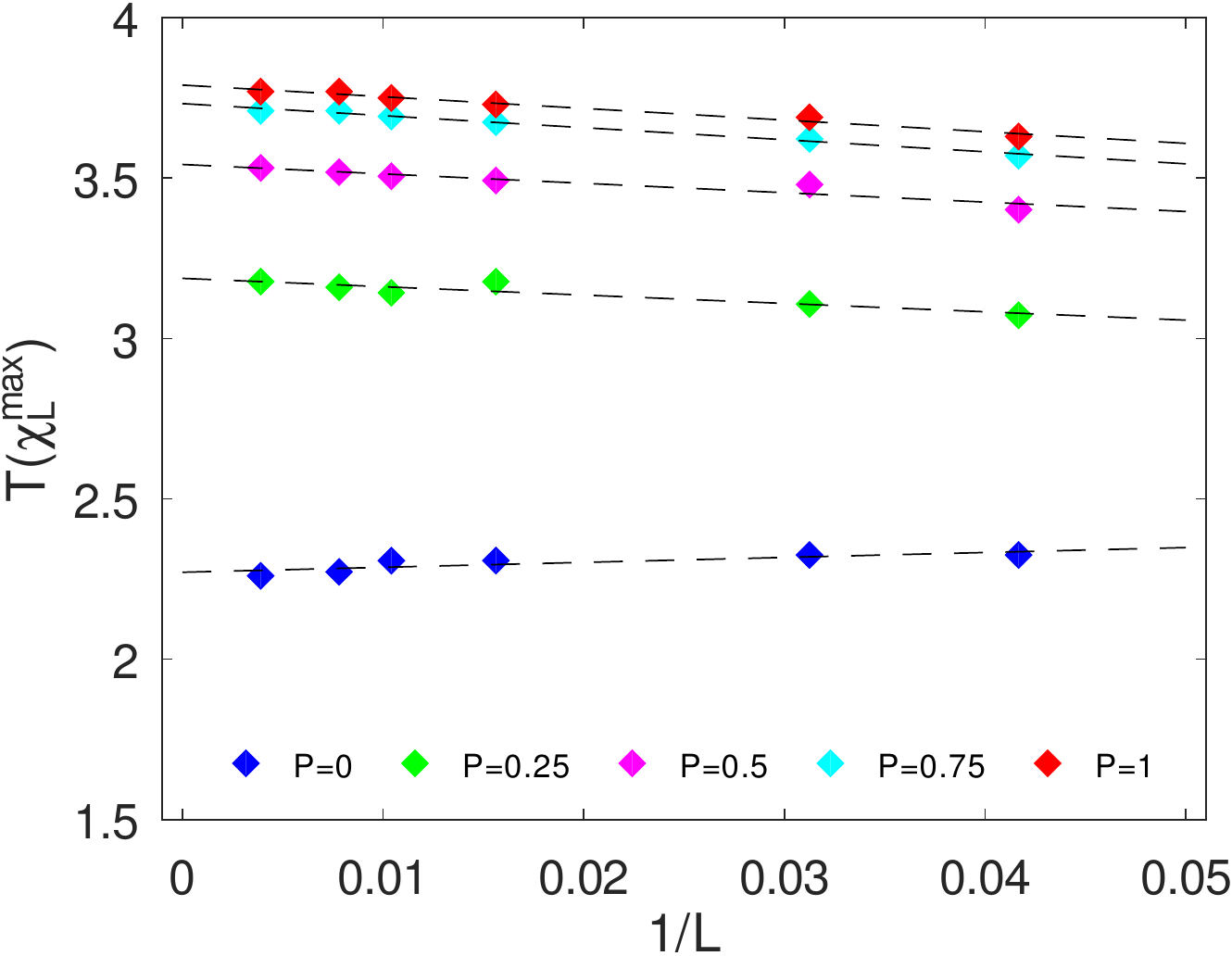}
\par\end{centering}
\caption{{\footnotesize{}Extrapolation of temperatures obtained by a lattice
with the linear size $24\protect\leq L\protect\leq256$ as $L\rightarrow\infty$,
for different values of $p$, as indicated in the figures. We have
found critical temperature $T_{c}(\infty)$ for $L\rightarrow\infty$
(see in Table \ref{tab:1}). The error bars are within the symbol
size.\label{fig:4}}}
\end{figure}

In Fig. \ref{fig:3}, we observed the same behavior of the finite-size
lattice of the magnetization $m_{L}$ (Fig. \ref{fig:3}(a)), the
fourth-order Binder cumulant $U_{L}$ (Fig. \ref{fig:3}(b)), and
the magnetic susceptibility $\chi_{L}$ (Fig. \ref{fig:3}(c)) as
a function of temperature $T$ for other values of $p$ ($0\leq p\leq1$).
In this case, we present only{\footnotesize{} }two different lattice
sizes $L=256$ and $L=24$. Together with the values of $p$ in the
A-SWN regime ($0<p\leq1$), we also calculate $p=0$, which is a very
know result of the two-dimensional Ising model in the regular lattice,
and it is calculated here by comparison with the different critical
behaviors.

The infinite-size extrapolation of the magnetic susceptibility peaks
at the critical point can be seen in Fig. \ref{fig:4} for some values
$p$ selected here. The values of $T_{c}(\infty)$ obtained by extrapolation
of $T(\chi_{L}^{max})$ for the linear size $24\leq L\leq256$ with
$L\rightarrow\infty$ can be found in Table \ref{tab:1}, and the
$T_{c}^{U}$ calculated by the crossing of the $U_{L}$ curves can
be seen in Table \ref{tab:2}, both for the different values of $p$.
In relation to $T_{c}$, we can see an agreement with the two methods
utilized here. When we increase the additive probability $p$, also
increase the $T_{c}$ of the system due to addiction of long-range
interaction $J_{ik}$ to the system and this consequently increase
the mean coordination number of the A-SWN. With $p=1$ all the sites
have one long-range interaction $J_{ik}$, therefore, the coordination
number $z$ is the same obtained in the SW-model studied in reference
\citep{19}. We can see that the critical temperature obtained here
$T_{c}(\infty)=3.79\pm0.02$ agrees with that obtained in the Ref.
\citep{19}, showing that the sublattices do not change the critical
behavior of the system and it is subjected to the same random long-range
interaction $J_{ik}$.

\begin{figure}
\begin{centering}
\includegraphics[clip,scale=0.55]{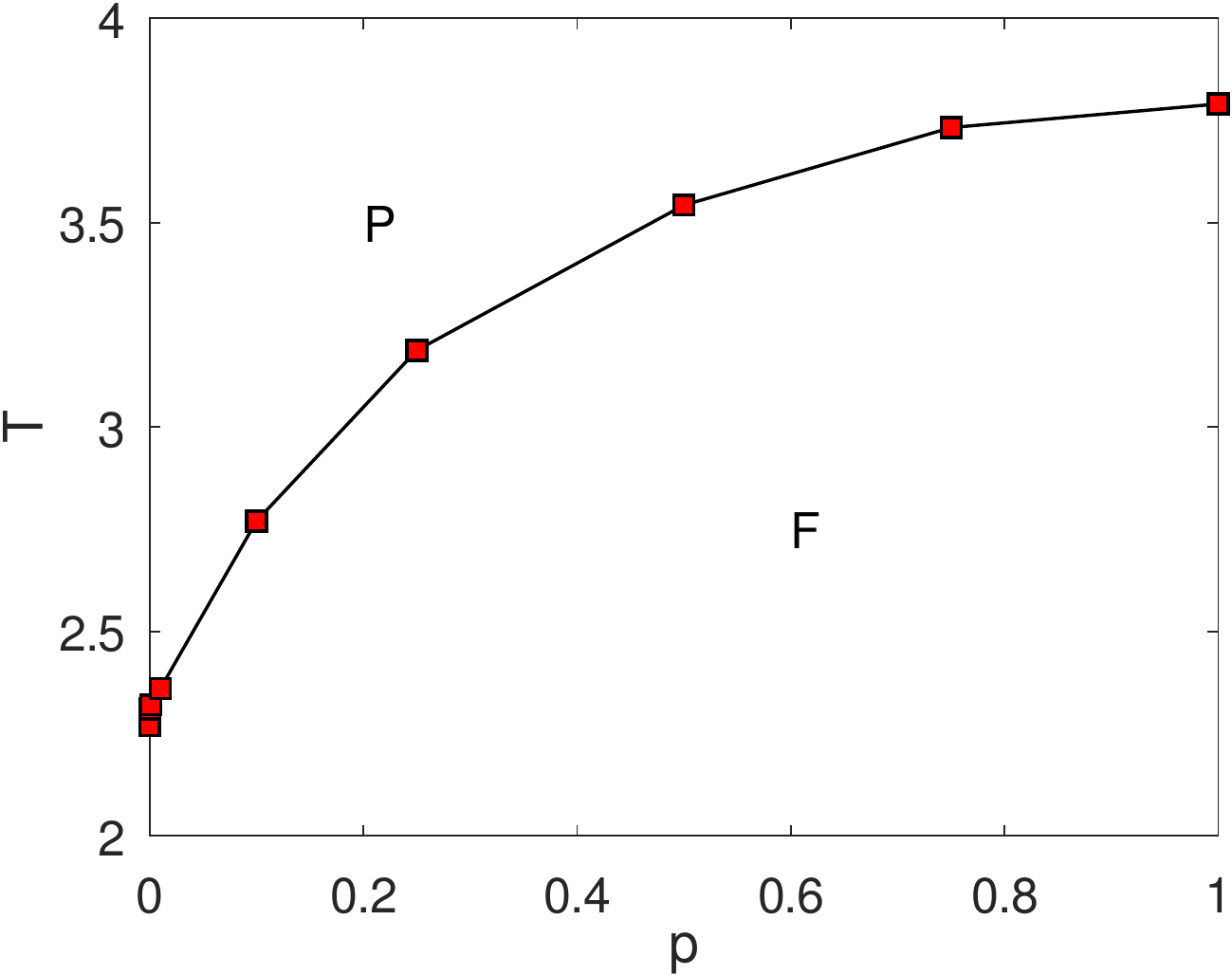}
\par\end{centering}
\caption{{\footnotesize{}Phases diagram of the Ising model on a 2D A-SWN, in
the temperature $T$ vs additive probability $p$ plane. $F$ and
$P$ are the ferromagnetic and paramagnetic phases, respectively.
The red-square dots represent second-order phase transition points
and full black line is a guide to the eyes. The error bars are within
the symbol size.\label{fig:5}}}
\end{figure}

We also have constructed the phases diagram showing the behavior of
temperature $T$ as a function of the addition probability $p$, and
can be seen in Fig. \ref{fig:5}. The phase diagram presents the $F$
ferromagnetic and $P$ paramagnetic phases and the full black line
represents a second-order transition line. We can see the critical
temperature increases with the addition of $J_{ik}$, beginning from
the standard Ising model in $p=0$, and increasing logaritmicaly in
order to $T\sim\ln(p^{0.45})$ for $p\gtrsim0.1$ until $p=1$.

\begin{figure}
\begin{centering}
\includegraphics[clip,scale=0.55]{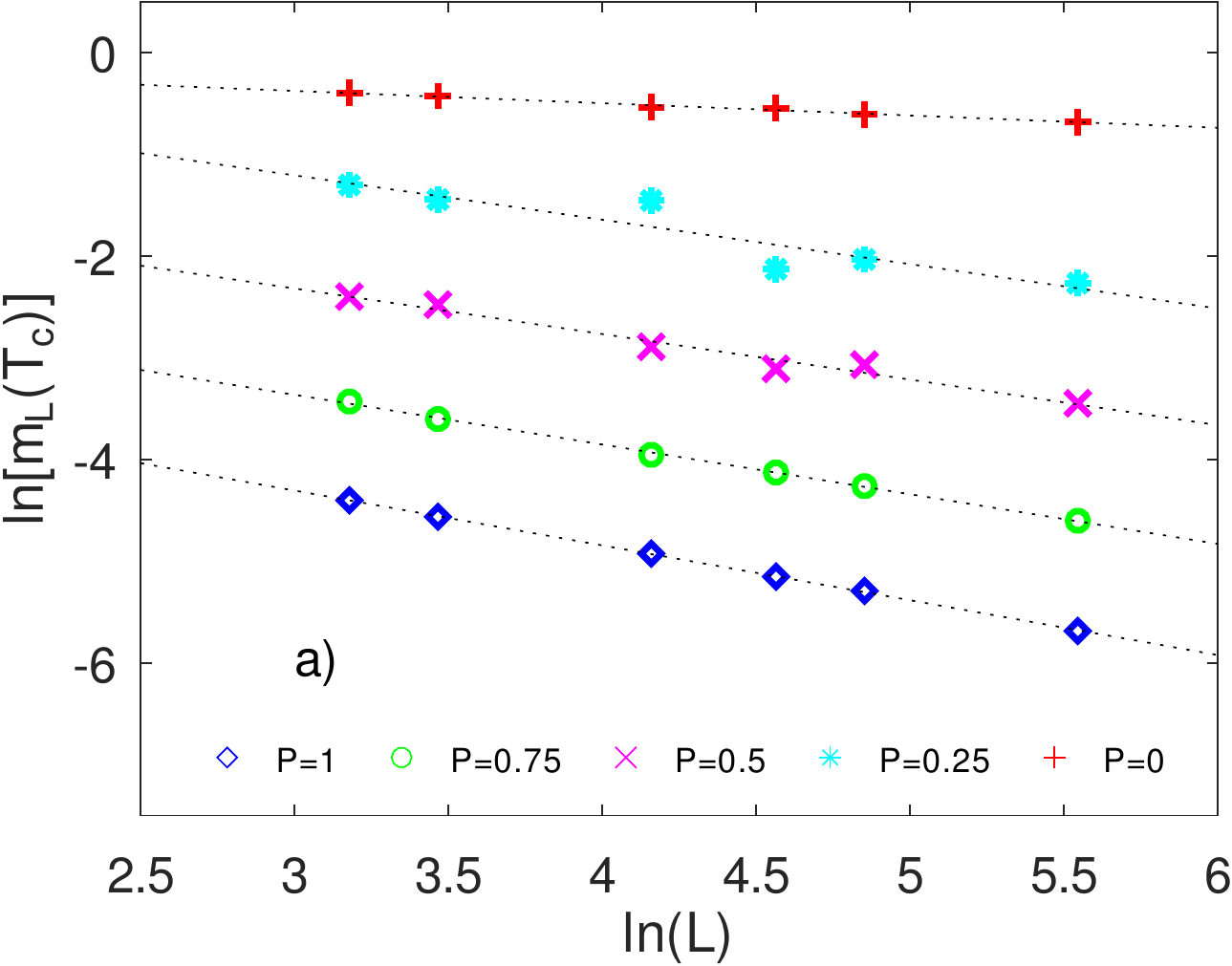}
\par\end{centering}
\begin{centering}
\includegraphics[clip,scale=0.55]{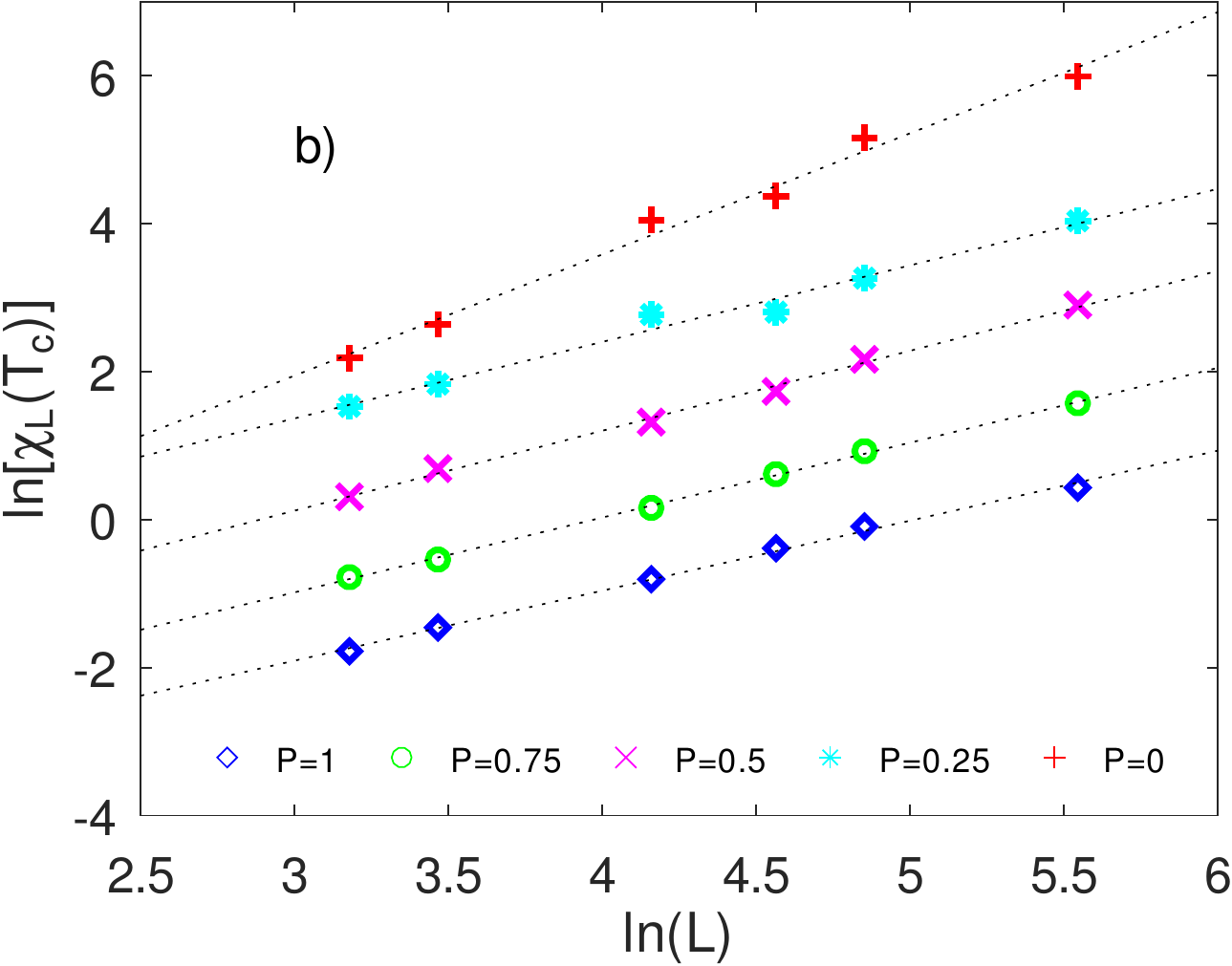}
\par\end{centering}
\begin{centering}
\includegraphics[clip,scale=0.55]{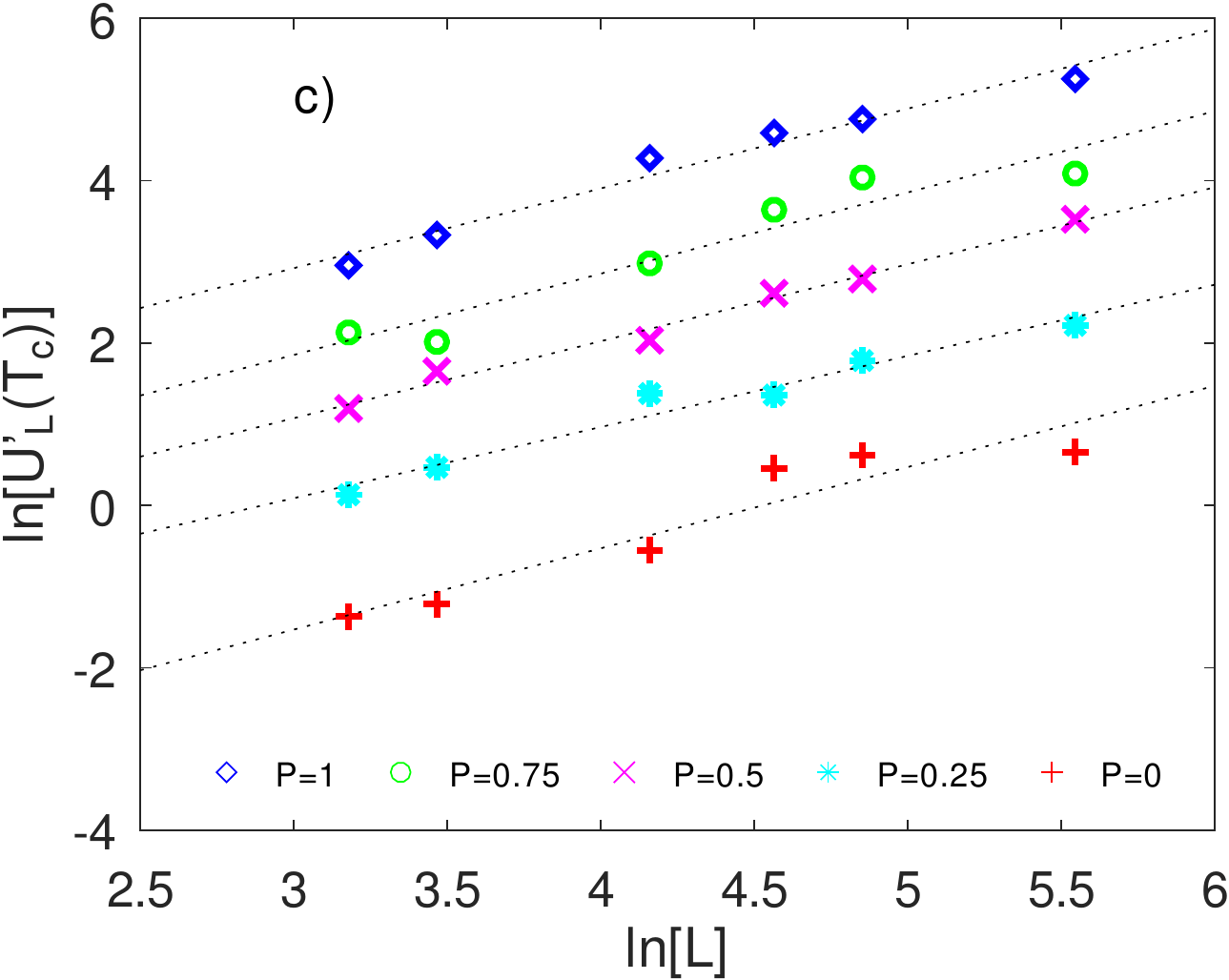}
\par\end{centering}
\caption{{\footnotesize{}The log-log plots of a) $m_{L}(T_{c})$, b) $\chi_{L}(T_{c})$,
and c) $U_{L}(T_{c})$ vs $L$, at the critical point and for different
values of $p$, as shown in the figures. The dotted lines are the
best fit for the data points. From these slopes, we have obtained
the critical exponents $\beta/\nu$, $\gamma/\nu$, and $\nu$ as
can see in Table \ref{tab:1}. The error bars are within the symbol
size.\label{fig:6}}}
\end{figure}

The critical point and the critical exponents for a phase transition
can be determined using the FSS analysis. We calculated the critical
exponents of the system by the slope of the best fit in the log-log
plot using the scaling relations Eq. (\ref{eq:7}), (\ref{eq:8}),
and (\ref{eq:10}). From the slope of the log-log plot for the magnetization
$m_{L}$ at the critical point for the different lattice sizes $L$
as indicated in Eq. (\ref{eq:7}), we found the ratio $-\beta/\nu$,
and can be seen in Fig. \ref{fig:6}(a). In the same way, the slope
of the log-log plot of the Eq. (\ref{eq:8}) give us the relation
$\gamma/\nu$ (see Fig. \ref{fig:6}(b)), and for the $\nu$ exponent
related to the correlation length of the system, we used the derivative
of the cumulant of the scaling relations Eq. (\ref{eq:10}), where
its slope in the log-log plot give us the relation $1/\nu$, see Fig.
\ref{fig:6}(c) \citep{20,21,22}. The log-log plot for scaling relations
for the some select values of $p$ can be seen in the Fig. \ref{fig:6},
which are in the A-SWN regime and for $p=0$ in contrast to the critical
behavior in a regular lattice and the A-SWN. As our interest is in
the slope of the log-log plot, we changed the linear coefficients
of the straight lines to separate the lines and thus making it easier
for the reader to see the fits.

We also have employed another procedure such that a family of curves
$m_{L}(T)$ and $\chi_{L}(T)$ collapse onto a single curve, the scaling
functions $m_{0}(L^{1/\nu}\varepsilon)$ and $\chi_{0}(L^{1/\nu}\varepsilon)$
respectively, as well as possible \citep{20,21,22}. In this procedure,
the best fitting can be obtained by adjusting the critical exponents
in the log-log plot of the isolated scaling functions as a function
of its variable $L^{1/\nu}\varepsilon$. Thus the critical exponents
that best collapse the curves are the possible critical exponents
of the system, having the verification of the exponent $\nu_{m}$
and $\nu_{\chi}$ (ver isso) by the magnetization and magnetic susceptibility
collapsed curves, respectively. For the values of $T$, we have $\varepsilon>0$
and $\varepsilon<0$, resulting in two curves on the data collapse,
where the best collapse should be near critical point and for the
largest values of $L$, because is in this region that the scale relations
are defined.

\begin{figure}
\begin{centering}
\includegraphics[clip,scale=0.55]{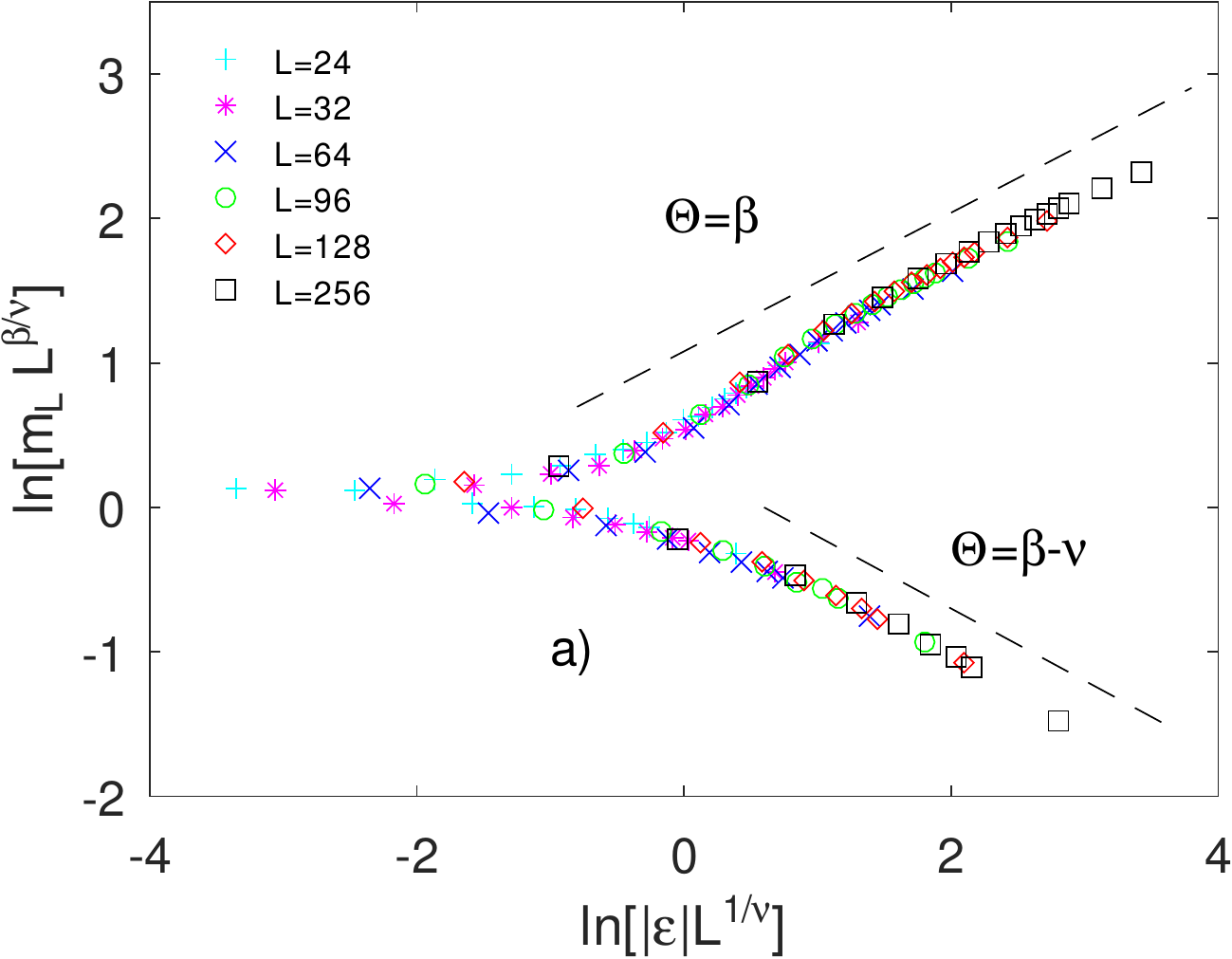}
\par\end{centering}
\begin{centering}
\includegraphics[clip,scale=0.55]{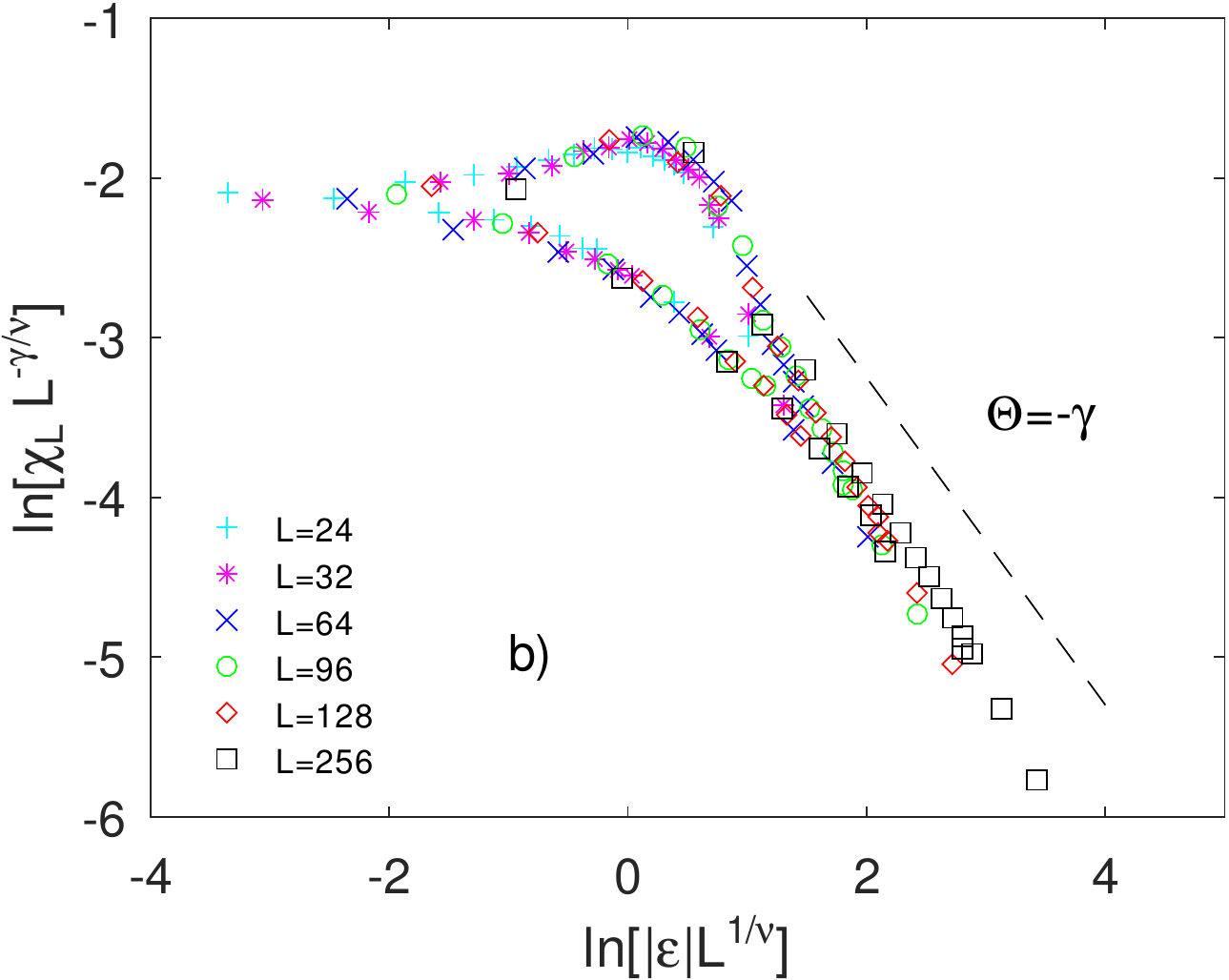}
\par\end{centering}
\caption{{\footnotesize{}Finite-size scaling (full data collapse) for the a)
magnetization $m_{L}$ and b) magnetic susceptibility $\chi_{L}$
for different values of $L$ as indicated in the figures. The parameter
$\varepsilon$ is defined by $\varepsilon=(T-T_{c})/T_{c}$. The dashed
lines represent the asymptotic behavior of the scaling functions.
The optimal values obtained for the critical exponents $\beta$, $\gamma$,
$\nu_{m}$, and $\nu_{\chi}$ can see in Table \ref{tab:2} for the
case $p=0.75$. The error bars are within the symbol size.\label{fig:7}}}
\end{figure}

\begin{figure}
\begin{centering}
\includegraphics[clip,scale=0.55]{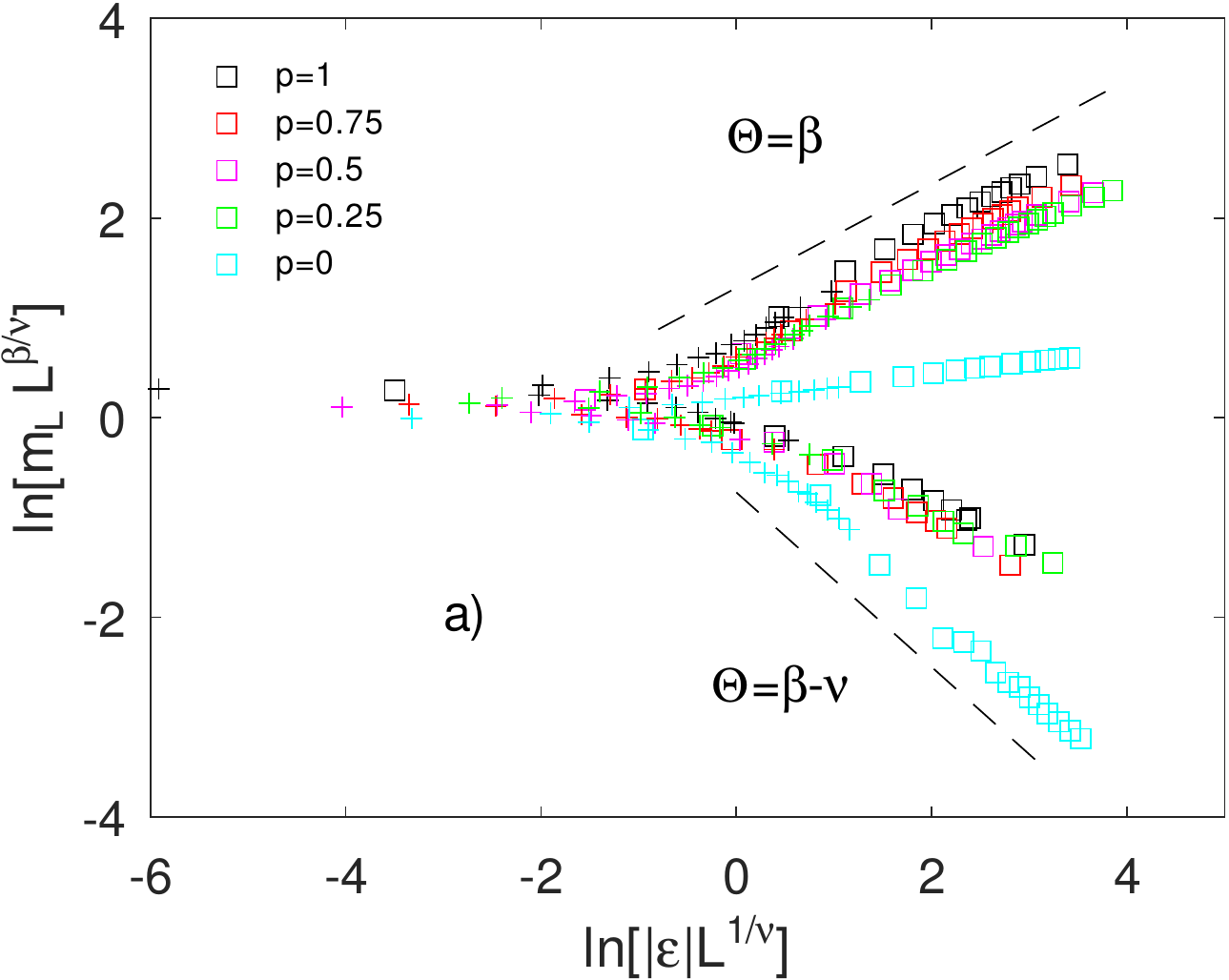}
\par\end{centering}
\begin{centering}
\includegraphics[clip,scale=0.55]{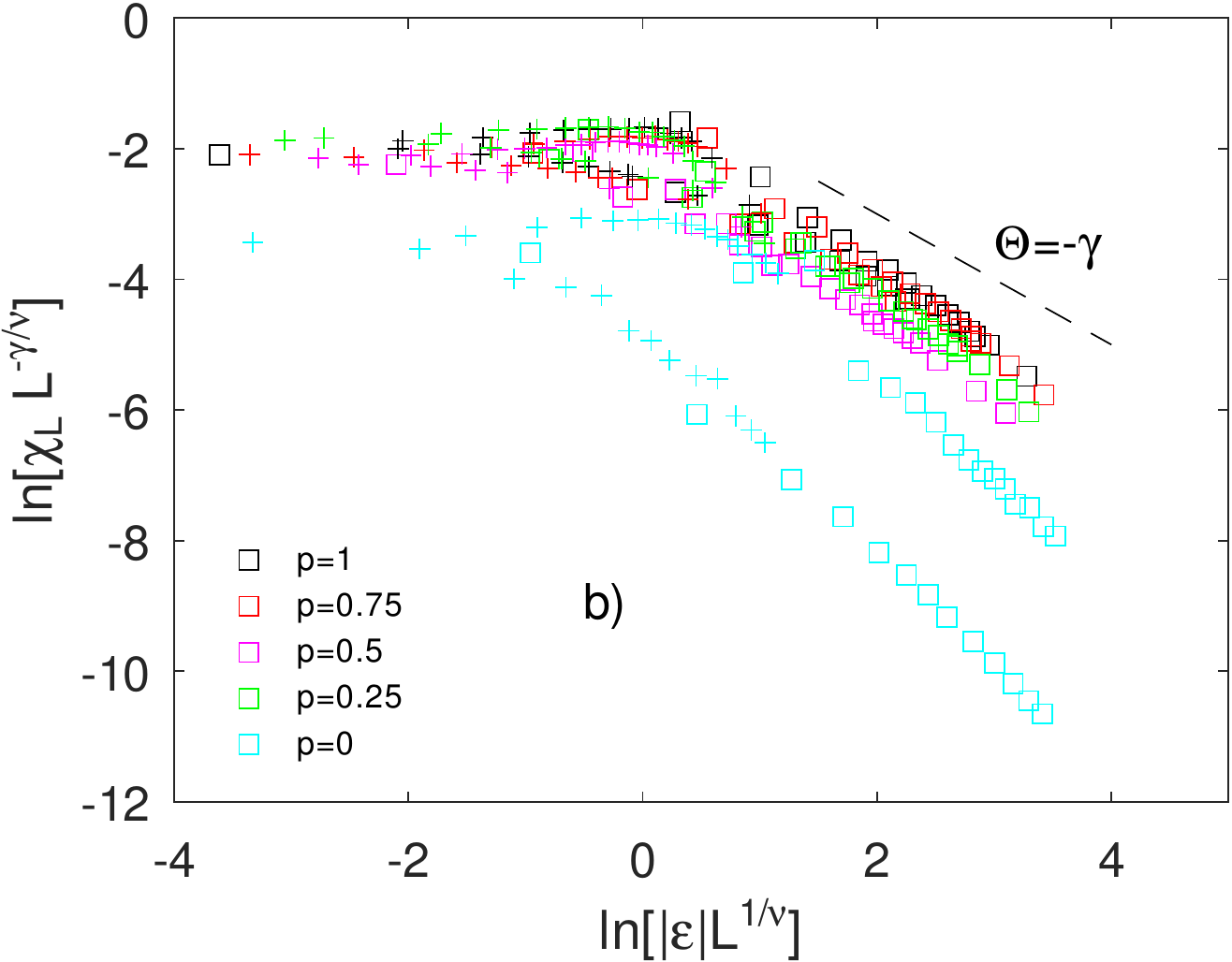}
\par\end{centering}
\caption{{\footnotesize{}Finite-size scaling (full data collapse) for the a)
magnetization $m_{L}$ and b) magnetic susceptibility $\chi_{L}$
for different values of $p$ as indicated in the figures. The parameter
$\varepsilon$ is defined by $\varepsilon=(T-T_{c})/T_{c}$. The dashed
lines represent the asymptotic behavior of the scaling functions.
The critical exponents used here are presented in Table \ref{tab:2}
and they were obtained using the best data collapse for all the lattice
sizes. Here, we exhibit only the lattice size $L=24\:(+)$ and $L=256\:(\square)$.
The error bars are within the symbol size.\label{fig:8}}}
\end{figure}

The best data collapse was obtained for $p=0.75$, and they are present
in Fig. \ref{fig:7}. In this figure, we have displayed the collapsed
curve for the magnetization $m_{L}(T)$ (Fig. \ref{fig:7}(a)) and
for the magnetic susceptibility $\chi_{L}(T)$ (Fig. \ref{fig:7}(b)),
and in both figures we have two curves representing the data with
$\varepsilon<0$ and $\varepsilon>0$. The data magnetization for
$\varepsilon<0$ corresponds to the slope $\Theta=\beta$, and the
data for $\varepsilon>0$ we have the slope $\Theta=\beta-\nu$. On
the other hand, for the magnetic susceptibility in both cases we have
the slope $\Theta=-\gamma$, being the superior curve referent to
$T<T_{c}$ and the inferior curve are data for $T>T_{c}$. For the
other values of $p$, we have obtained the best-fitting and the collapsed
curves for the lattice sizes $L=24$ and $L=256$, which can be seen
in Fig. \ref{fig:8}(a) for $m_{L}(T)$ and in Fig. \ref{fig:8}(b)
for $\chi_{L}(T)$. 

The results for critical exponents obtained by the best fit of the
log-log plots of thermodynamic quantities near $T_{c}$ are exhibited
in Table \ref{tab:1} and the critical exponents based on the data
collapses are shown in Table \ref{tab:2}. In the both methods we
found equivalent values for the critical exponent for each $p$, being
the best results here based on the data collapse, i.e., due to the
fact that the values are approximately closer to the values of the
Ising model in the regular lattice. When $p=0$, we have the known
critical exponents of the 2D Ising model (regular square lattice)
for exact solutions and Monte Carlo simulations \citep{24} and given
by $\beta=1/8$, $\gamma=7/4$ and $\nu=1$. On the other hand, when
we increase the addition probability $p$, the $J_{ik}$ change these
critical exponents until to reach the mean-field behavior as observed
in Ref. \citep{19}, $\beta=1/2$, $\gamma=1$ and $\nu=1/2$ (using
$L^{2}=N$ in the scaling relations instead of only $L$). We also
can observe a smooth variance of these exponents along the A-SWN regime,
tending to the regular lattice Ising model exponents as the $p$ decrease. 

We know that the critical exponents are not independent of each other,
but related by simple scaling relations. The scaling relation well-called
is the hyperscaling relation $d=2\beta/\nu+\gamma/\nu$ (scaling laws
in which the spatial dimension $d$ appears explicitly) and which
give us the spatial dimension $d$ of the system \citep{24}. Therefore,
with the hyperscaling relation, we can see that the system has approximately
the same critical exponents in the A-SWN regime (see Table \ref{tab:1}
and \ref{tab:2} for $0<p\leq1$) and its spatial dimension is $d\cong2.0$.
The universality class can be defined as the complete set of exponents
at the phase transition, as in the case of the second-order phase
transition, where very different systems from each other can share
the same set of critical exponents. In general, these systems share
the same spatial dimensions, symmetries, and range interactions. Thus,
here we have a system with the same symmetries (up-down) and spatial
dimensionality $d=2$ but the range of interactions can be different
by adding the long-range interactions $J_{ik}$. Therefore, here we
have a system with a set of critical exponents, and consequently indicating
a universality class from the mean-field critical exponents of Ising-like
systems, for the A-SWN regime ($0<p\leq1)$.

\begin{table}[H]
\caption{{\footnotesize{}Critical exponents obtained by the best fit of the
log-log plots of termodynamic quantities near $T_{c}$ (see Fig. \ref{fig:6}),
$T_{c}(\infty)$ based in the values of the infinite-lattice critical
temperature (see Fig. \ref{fig:4}) and for the different values of
$p$. \label{tab:1}}}

\centering{}%
\begin{tabular}{c|c|c|c|c}
\hline 
$p$ & $\beta/\nu$ & $\gamma/\nu$ & $\nu$ & $T_{c}(\infty)$\tabularnewline
\hline 
\hline 
$0$ & $0.120\pm0.007$ & $1.64\pm0.08$ & $1.00\pm0.09$ & $2.27\pm0.03$\tabularnewline
\hline 
$0.25$ & $0.44\pm0.08$ & $1.03\pm0.06$ & $1.14\pm0.08$ & $3.19\pm0.04$\tabularnewline
\hline 
$0.5$ & $0.45\pm0.03$ & $1.08\pm0.03$ & $1.05\pm0.06$ & $3.54\pm0.04$\tabularnewline
\hline 
$0.75$ & $0.49\pm0.01$ & $1.01\pm0.02$ & $1.00\pm0.09$ & $3.73\pm0.01$\tabularnewline
\hline 
$1$ & $0.54\pm0.01$ & $0.95\pm0.02$ & $1.02\pm0.08$ & $3.79\pm0.02$\tabularnewline
\hline 
\end{tabular}
\end{table}

\begin{table*}[t]
\caption{{\footnotesize{}Critical exponents obtained by the best data collapse
(see Fig. \ref{fig:7} and Fig. \ref{fig:8}), $T_{c}^{U}$ based
on the crossing of the $U_{L}$ curves for different lattice sizes
$L$ and for the several values of $p$. The spatial dimension is
calculated by hyperscaling relation $d=2\beta/\nu+\gamma/\nu$.\label{tab:2}}}

\centering{}%
\begin{tabular}{c|c|c|c|c|c|c}
\hline 
$p$ & $\beta$ & $\gamma$ & $\nu_{m}$ & $\nu_{\chi}$ & $T_{c}^{U}$ & $d$\tabularnewline
\hline 
\hline 
$0$ & $0.125\pm0.002$ & $1.75\pm0.05$ & $1.00\pm0.05$ & $1.00\pm0.05$ & $2.27\pm0.01$ & $2.0$\tabularnewline
\hline 
$0.25$ & $0.44\pm0.04$ & $1.12\pm0.05$ & $0.95\pm0.05$ & $1.05\pm0.09$ & $3.18\pm0.03$ & $2.0$\tabularnewline
\hline 
$0.5$ & $0.45\pm0.03$ & $1.15\pm0.03$ & $0.95\pm0.05$ & $1.05\pm0.03$ & $3.54\pm0.02$ & $2.05$\tabularnewline
\hline 
$0.75$ & $0.48\pm0.02$ & $1.02\pm0.03$ & $0.98\pm0.04$ & $0.98\pm0.04$ & $3.73\pm0.02$ & $2.02$\tabularnewline
\hline 
$1$ & $0.52\pm0.03$ & $1.00\pm0.04$ & $0.98\pm0.03$ & $1.00\pm0.03$ & $3.79\pm0.01$ & $2.06$\tabularnewline
\hline 
\end{tabular}
\end{table*}

\section{Conclusions\label{sec:Conclusions}}

In this work, we have developed the Monte Carlo simulations to study
of thermodynamic quantities and critical behavior of the Ising model
on a 2D A-SWN. With the thermodynamic quantities and the fourth-order
Binder cumulant we have obtained the critical point of the second-order
phase transitions in the A-SWN regime $(0<p\leq1)$. The ordered to
disordered phase transitions result in a phase diagram from the ferromagnetic
$F$ to paramagnetic $P$ phase transitions, in which we can observe
an increase in the critical temperature $T_{c}$ of the system as
the addition of the long-range interactions $J_{ik}$. Through Monte
Carlo simulations and finite-size scaling arguments, we calculated
the static critical exponents $\beta$, $\gamma$, and $\nu$, and
we concluded that based on the critical exponents, this model is in
the same universality class of the pure Ising model in two dimensions
on mean-field approximation in the A-SWN regime $(0<p\leq1)$. However,
we also have obtained the critical exponents for the system in the
case $p=0$, and we have observed that the system has a critical behavior
from the regular square lattice Ising model with critical exponents
calculated using exact calculation and Monte Carlo simulation. Our
results, of the change in $T_{c}$ and the mean-field behavior is
in agreement with the observed behavior of disorder with shortcuts
added to the Ising model in R-SWN and A-SWN \citep{5,14,16,17,19}.
Therefore, the direction in the long-range interactions between sublattices
of the network, do not change the SWN behavior, being that the Ising
Model on a 2D A-SWN scales logarithmically as a function of $p$ to
the mean-field critical behavior.

\begin{acknowledgments}
This work was partially supported by the Brazilian agencies CNPq,
UFMT and FAPEMAT.
\end{acknowledgments}

\end{document}